\documentclass[12pt]{iopart}

\usepackage{units}
\usepackage{graphicx}
\usepackage{booktabs}
\usepackage{subeqnarray}
\usepackage{cite}
\usepackage{epstopdf}
\usepackage{citesort}
\usepackage[american]{babel}
\usepackage{graphicx}
\expandafter\let\csname equation*\endcsname\relax
\expandafter\let\csname endequation*\endcsname\relax
\usepackage[intlimits]{amsmath}
\usepackage{braket}
\usepackage{bm}
\usepackage{amssymb}
\usepackage{epstopdf}

\begin{document}
\title[Covariant Description of Nonlinear Transformation Optics]{Covariant Description of Transformation Optics in Linear and Nonlinear Media}
\author{Oliver Paul$^{1}$, Marco Rahm$^{1,2}$}
\address{$^1$Department of Physics and Research Center OPTIMAS, University of Kaiserslautern, Germany \\
\vspace{5 pt} $^2$Fraunhofer Institute for Physical Measurement Techniques IPM, Freiburg, Germany}

\date{\today}

\begin{abstract}
The technique of transformation optics (TO) is an elegant method for the design of electromagnetic
media with tailored optical properties. In this paper, we focus on the formal structure of TO
theory. By using a complete covariant formalism, we present a general transformation law that holds
for arbitrary materials including bianisotropic, magneto-optical, nonlinear and moving media. 
Due to the
principle of general covariance, the formalism is applicable to arbitrary space-time coordinate transformations and automatically
accounts for magneto-electric coupling terms.
The formalism is demonstrated for the
calculation of the second harmonic generation in a twisted TO concentrator.
\end{abstract}



\section{Introduction}
The field of transformation optics (TO) has drawn a lot of scientific interest in the last few years~\cite{pendry2006,leonhardt2006,milton2006,cummer2007,kildishev2008,rahm2008}. 
By this design methodology, the form-invariance of Maxwell's equations under coordinate transformation is used to tailor the optical properties of an electrodynamic medium. The majorities of TO applications focus on the design of the linear material parameters. 
In that case, a coordinate transformation is used
to engineer the linear constitutive parameters of a medium such that the wave trajectory follows a desired path~\cite{kundtz_smith_ieee10}.
An interesting application of this concept is the realization of
electromagnetic invisibility cloaks~\cite{schurig2006,cai2007,landy2010,leonhardt2011,urzhumov2011}---devices in which light is guided around a certain region of space rendering the interior of the region invisible for an external observer. 
Cloaking devices belong to the most prominent applications in which the TO concept is successfully applied for designing the linear material properties of a medium and have been extensively reported in 
the literature~\cite{novitsky2009,kwon2008,rahm2008b,han2010,nicolet2008,wang_zhou10,jiang2008}.

In contrast, only little work in the research of TO media addresses the transformation of nonlinear material properties. 
Media with nonlinear response, however, provide a number of interesting effects including sum- and difference-frequency generation, parametric amplification and oscillation, stimulated scattering, self-phase modulation and self-focusing~\cite{kelley1965,faust1966,bloembergen1967,manassah1988,shen1989,agrawal1989,macklin1993,myers1995,dubietis2006} and play a key role in modern optical technology~\cite{shen1976,evans1997,boyd2008}. 
Consequently, the extension of the TO concept to nonlinear media
is expected to offer a variety of new opportunities for engineering optical media and for the construction of novel electromagnetic devices~\cite{bergamin2011}. 
Apart from that, the TO concept also provides a promising calculation method in modeling complex nonlinear systems 
whenever it is possible to find a coordinate transformation such that the geometry of the system takes a much simpler form in the transformed space.

A basic prerequisite for a successful integration of nonlinear effects into the TO concept is a general transformation law for linear and nonlinear material parameters under space-time coordinate transformation. A first step in this direction is suggested in~\cite{bergamin2011} 
where the transformation of certain classes of nonlinear materials under purely spatial transformations is studied. 
The aim of the following paper is to generalize this approach and to provide a rigorous theoretical framework for arbitrary nonlinear materials and for both temporal and spatial coordinate transformations. The formalism is presented in a  manifestly covariant form that allows the simultaneous treatment of electric, magnetic and magneto-electric cross-coupling effects and encompasses all types of space-time transformations with a particular reference to moving media~\cite{post1997,thompson2010,thompson2011}.

The paper is organized as follows: In the first part, we introduce the used tensor notation and explain how the defined tensors are related to the common linear and nonlinear material parameters such as the permittivity, permeability or the quadratic electro-optic coefficients. 
Subsequently, we exploit the fundamental principle of relativity to derive a general transformation law for nonlinear media under continuous space-time coordinate transformations. 
In this context, a special emphasis is given to the transformation of nonlinear constitutive parameters in moving media and the class of time-independent, spatial transformations which play a central role for the design of TO devices. 
In the final part of the paper, we illustrate and explain the derived expressions by means of an explicit calculation example. 
By deriving the second harmonic generation in a nonlinear, twisted field concentrator, 
we show that an appropriate coordinate transformation can provide a significant alleviation in the formal treatment of a complex nonlinear problem and, thus, allows a convenient calculation of an otherwise sophisticated process.

\section{The covariant material equation}
In order to establish a common basis for the following discussion and to introduce the used notation, we start
with a short review of the covariant description of the electrodynamic theory. In this paper, all
quantities are expressed in Gaussian units. In this case, the Maxwell equations in matter take the
form:
\begin{alignat}{2}
\label{eq:maxwell_drei1}
\text{div}\,\boldsymbol{B} &= 0, &\quad \quad   \text{rot}\,\boldsymbol{E} + \frac{1}{c}\frac{\partial \boldsymbol{B}}{\partial t} & = 0, \\[2mm]
\label{eq:maxwell_drei2} \text{div}\,\boldsymbol{D} &=4\pi\rho, &\quad \quad
\text{rot}\,\boldsymbol{H}  - \frac{1}{c}\frac{\partial \boldsymbol{D}}{\partial t} & =
\frac{4\pi}{c}\boldsymbol{j}
\end{alignat}
where $\boldsymbol{E}$, $\boldsymbol{H}$, $\boldsymbol{D}$, $\boldsymbol{B}$, $\rho$,
$\boldsymbol{j}$ and $c$ denote the electric and magnetic field, the electric and magnetic flux
density, the charge, the current density and the speed of light in vacuum, respectively. By using
the four-current $(j^\nu)=(c\rho,\boldsymbol{j})$, the antisymmetric field strength tensor
\begin{align}
(F_{\mu\nu})=
\begin{pmatrix}
0 & E_x & E_y & E_z \\
-E_x & 0 & -B_z & B_ y\\
-E_y & B_z & 0 & -B_x \\
-E_z & -B_y & B_x & 0
\end{pmatrix}\label{eq:field_strength_tensor}
\end{align}
and the antisymmetric displacement tensor
\begin{align}
(\mathcal{D}_{\mu\nu})=
\begin{pmatrix}
0 & D_x & D_y & D_z \\
-D_x & 0 & -H_z & H_ y\\
-D_y & H_z & 0 & -H_x \\
-D_z & -H_y & H_x & 0
\end{pmatrix}\label{eq:displacement_tensor},
\end{align}
the Maxwell equations can be covariantly expressed as:
\begin{align}
\label{eq:maxwell_kov1}\partial^\mu F_{\nu\sigma} + \partial^\sigma{F}_{\mu\nu}  + \partial^\nu {F}_{\sigma\mu}  &= 0, \\
\label{eq:maxwell_kov2} \partial^\mu \mathcal{D}_{\mu\nu} &= \frac{4\pi}{c} j_\nu
\end{align}
where we use the Einstein summation convention\footnote[1]{Throughout this paper, Greek indices run from 0 to 3 while Latin indices run from 1 to 3.} and the contravariant derivation  $(\partial^\mu) =(\frac{1}{c}\partial_t,-\nabla)$.
In the covariant notation, (\ref{eq:maxwell_kov1}) contains the two homogeneous Maxwell
equations whereas (\ref{eq:maxwell_kov2}) contains the two inhomogeneous Maxwell equations.
Note that the signs used in the tensor notation above depend on the convention used for the metric
tensor. In this paper, we use the metric tensor given
by $\eta^{\mu\nu}=\text{diag}(1,-1,-1,-1)$. The relation between the tensors $F_{\mu\nu}$ and
$\mathcal{D}_{\mu\nu}$ is given by
\begin{align}
\label{eq:matgleichung} \mathcal{D}_{\mu\nu} = F_{\mu\nu} +4\pi \mathcal{P}_{\mu\nu}
\end{align}
with the polarization-magnetization tensor
\begin{align}
(\mathcal{P}_{\mu\nu})=
\begin{pmatrix}
0 & P_x & P_y & P_z \\
-P_x & 0 & M_z & -M_ y\\
-P_y & -M_z & 0 & M_x \\
-P_z & M_y & -M_x & 0
\end{pmatrix}.
\end{align}
Tensor equation~(\ref{eq:matgleichung}) is equivalent to the common material equations:
\begin{align}
\boldsymbol{D} &= \boldsymbol{E} + 4\pi\boldsymbol{P}, \nonumber\\
\boldsymbol{H} &= \boldsymbol{B} - 4\pi\boldsymbol{M}.
\end{align}
In many cases, the quantitative relation between the polarization-magnetization
tensor~$\mathcal{P}_{\mu\nu}$ and the field strength tensor~$F^{\mu\nu}$ can be expressed in a
power series according to:
\begin{align}
\nonumber
\mathcal{P}_{\mu\nu}  = &\chi_{\mu\nu}^{\sigma\kappa}\,F_{\sigma\kappa} + \chi_{\mu\nu}^{\sigma\kappa\alpha\beta}\,F_{\sigma\kappa}F_{\alpha\beta} 
+ \chi_{\mu\nu}^{\sigma\kappa\alpha\beta\gamma\delta}F_{\sigma\kappa}F_{\alpha\beta}F_{\gamma\delta} + \cdots \nonumber\\[2mm]
= &\sum_{n\ge1} \chi_{\mu\nu}^{\alpha_1\beta_1\cdots\alpha_n\beta_n}F_{\alpha_1\beta_1}\dots
F_{\alpha_n\beta_n}.\label{eq:polarisation}
\end{align}
Note that in this notation, the index pair $\mu\nu$ is fixed while the other indices are summed
over. From the antisymmetry of the tensors $F_{\mu\nu}$ and $\mathcal{P}_{\mu\nu}$ and the
commutativity of the products $F_{\alpha_i\beta_i}F_{\alpha_j\beta_j}$, it follows that the
coefficients~$\chi_{\mu\nu}^{\alpha_1\beta_1\cdots\alpha_n\beta_n}$ are antisymmetric under
exchange of~$\mu\leftrightarrow\nu$, symmetric under exchange of two pairs
$\alpha_i\beta_i\leftrightarrow\alpha_j\beta_j$ and antisymmetric under exchange of two indices
$\alpha_i\leftrightarrow\beta_i$ within one pair~$\alpha_i\beta_i$. For the quadratic term, this
means for example:
\begin{align}
\chi_{\mu\nu}^{\alpha_1\beta_1\alpha_2\beta_2} = -\chi_{\nu\mu}^{\alpha_1\beta_1\alpha_2\beta_2} =
\chi_{\mu\nu}^{\alpha_2\beta_2\alpha_1\beta_1} = -\chi_{\mu\nu}^{\beta_1\alpha_1\alpha_2\beta_2}.
\label{eq:symmetries}
\end{align}
It is obvious that additional, inherent symmetries (such as the Kleinman symmetry or spatial symmetries
given by the point symmetry class of the medium) further reduce the number of independent
coefficients. However, in order to provide a general description, we only consider the basic
symmetries given in (\ref{eq:symmetries}). Inserting (\ref{eq:polarisation}) into the
material equation~(\ref{eq:matgleichung}) yields:
\begin{align}
\mathcal{D}_{\mu\nu} & = F_{\mu\nu} +4\pi \mathcal{P}_{\mu\nu} \nonumber\\[2mm]
& = F_{\mu\nu} + 4\pi\sum_{n\ge1} \chi_{\mu\nu}^{\alpha_1\beta_1\cdots\alpha_n\beta_n}F_{\alpha_1\beta_1}\dots F_{\alpha_n\beta_n}\nonumber\\
& =  4\pi\sum_{n\ge1} \chi_{\mu\nu}^{\alpha_1\beta_1\cdots\alpha_n\beta_n}F_{\alpha_1\beta_1}\dots
F_{\alpha_n\beta_n}\label{eq:matgleichung2}
\end{align}
where in the last line, we performed a re-definition of the linear coefficient
$\chi_{\mu\nu}^{\alpha_1\beta_1}$ to include the free-space space contribution (see appendix).
\subsection{The linear term}
To become familiar with the tensor notation, it is instructive to first consider only the linear term
on the right-hand side of (\ref{eq:matgleichung2}):
\begin{align}
\label{eq:linearterm} \mathcal{D}_{\mu\nu} & = 4\pi\chi_{\mu\nu}^{\sigma\kappa}\,F_{\sigma\kappa}.
\end{align}
By using the symmetry properties of $\chi_{\mu\nu}^{\sigma\kappa}$, it follows for the components
of the electric displacement field~$\boldsymbol{D}$:
\begin{align}
D_i&=\mathcal{D}_{0i} =4\pi\chi_{0i}^{\sigma\kappa}\,F_{\sigma\kappa} =
8\pi\chi_{0i}^{0j}\,F_{0j}+4\pi\chi_{0i}^{kl}\,F_{kl} \nonumber\\
&=\epsilon_i^j E_j + \xi_i^jB_j \label{eq:d-feld}
\end{align}
where $\epsilon_i^j$ denotes the permittivity tensor and $\xi_i^j$ is the tensor of the electro-magnetic coupling (remember, Latin indices run from 1 to 3). The last line follows from the fact that $F_{0j}$ addresses the components of the $\boldsymbol{E}$-field while $F_{kl}$ addresses the components of the $\boldsymbol{B}$-field.
In a similar way, one
finds for the components of the $\boldsymbol{H}$-field:
\begin{align}
H_i &= -\frac{1}{2}g_i^{mn}\mathcal{D}_{mn} = -2\pi
g_i^{mn}\chi_{mn}^{\sigma\kappa}\,F_{\sigma\kappa}\nonumber\\
&= \zeta_i^jE_j + (\mu^{-1})_i^jB_j \label{eq:h-feld}
\end{align}
with the totally antisymmetric Levi-civita tensor $g_i^{mn}$ (with $g_1^{23}=1$), the
permeability~$\mu_i^j$ and the magneto-electric coupling tensor~$\zeta_i^j$. In summary, the linear tensor equation $\mathcal{D}_{\mu\nu} =
4\pi\chi_{\mu\nu}^{\sigma\kappa}\,F_{\sigma\kappa}$ can be expressed in the common three-formalism
as:
\begin{align}
\begin{pmatrix}
\boldsymbol{D}\\
\boldsymbol{H}
\end{pmatrix}
=
\begin{pmatrix}
\epsilon & \xi \\
\zeta & \mu^{-1}
\end{pmatrix}
\begin{pmatrix}
\boldsymbol{E}\\
\boldsymbol{B}
\end{pmatrix}.
\end{align}
This is the general equation of a linear bianisotropic medium. The relations between the parameters
$\epsilon$, $\mu$, $\xi$ and $\zeta$ and the four-tensor $\chi^{\sigma \kappa}_{\mu\nu}$ can be
derived by comparing the coefficients occurring in (\ref{eq:d-feld}) and (\ref{eq:h-feld}). As shown
in the appendix, one finds:
\begin{alignat}{2}
\epsilon_i^j &= 8\pi\chi_{0i}^{0j}, & \quad\quad \xi_i^j &= -4\pi g_{mn}^j \chi_{0i}^{mn},\nonumber\\
\zeta_i^j &= -4\pi g_i^{mn}\chi^{0j}_{mn}, & \quad\quad (\mu^{-1})_i^j &= 2\pi g^j_{mn}
g_i^{kl}\chi_{kl}^{mn}.\label{eq:lineare_terme}
\end{alignat}
\subsection{Nonlinear terms}
Next, we consider quadratic contributions to the polarization-magnetization
tensor~$\mathcal{P}_{\mu\nu}$, i.e.~contributions that depend on the product of two components of
the field strength tensor. These are given by the second summand in (\ref{eq:polarisation}) and
have the form:
\begin{align}
\nonumber \mathcal{P}^{(2)}_{\mu\nu} & =
\chi_{\mu\nu}^{\sigma\kappa\alpha\beta}\,F_{\sigma\kappa}F_{\alpha\beta}.
\end{align}
For simplicity, we restrict to the components of the second-order electric polarization
$P^{(2)}_i=\mathcal{P}^{(2)}_{0i} $ (a similar derivation applies for the magnetization). As for
the linear term, we can use the symmetry properties of~$\chi_{\mu\nu}^{\sigma\kappa\alpha\beta}$ to
split the summation into three terms:
\begin{align}
\nonumber
P^{(2)}_i & =  \chi_{0i}^{\sigma\kappa\alpha\beta}\,F_{\sigma\kappa}F_{\alpha\beta}\\[2mm]
 & = 4\chi_{0i}^{0k0m}\,F_{0k}F_{0m} + 4 \chi_{0i}^{0kmn}\,F_{0k}F_{mn} + \chi_{0i}^{klmn}\,F_{kl}F_{mn}
 \nonumber\\[2mm]
 & = \underbrace{a_i^{jk}E_jE_k}_\text{Pockels effect, multi-wave mixing} + \underbrace{b_i^{jk}E_jB_k}_\text{Faraday effect} + \enspace c_i^{jk}B_jB_k  \label{eq:zweite_ordnung}
\end{align}
where we have introduced the nonlinear material coefficients~$a_i^{jk}$, $b_i^{jk}$, $c_i^{jk}$. In
other words, the expression $\mathcal{P}^{(2)}_{\mu\nu}  =
\chi_{\mu\nu}^{\sigma\kappa\alpha\beta}\,F_{\sigma\kappa}F_{\alpha\beta}$ describes all second-order electric, magnetic and magneto-electric cross-coupling effects in a single equation. 
Similarly, one finds for the nonlinear polarization
of the third order:
\begin{align}
P^{(3)}_i & =  \chi_{0i}^{\sigma\kappa\alpha\beta\mu\nu}\,F_{\sigma\kappa}F_{\alpha\beta}F_{\mu\nu}\nonumber\\[2mm]
  & = \underbrace{a_i^{jkl}E_jE_kE_l}_\text{Kerr effect} \enspace  + \enspace  b_i^{jkl}E_jE_kB_l  \enspace +   \underbrace{c_i^{jkl}E_jB_kB_l}_\text{Cotton-Mouton effect}   + \enspace  d_i^{jkl}B_jB_kB_l \label{eq:dritte_ordnung}
\end{align}
and so on. Note: while the quadratic contributions in (\ref{eq:zweite_ordnung}) contain $6+9+6=21$
independent coefficients, the cubic contributions in (\ref{eq:dritte_ordnung}) contain already
$10+18+18+10=56$ independent coefficients (see appendix).

\section{Coordinate transformation}
The particular advantage of the covariant formulation is its form-invariance under coordinate transformations
\begin{align}
x^\alpha \longrightarrow x^{\alpha'}(x^\alpha)
\end{align}
where $(x^\alpha)=(ct,\boldsymbol{r})$ is the coordinate vector of the space time. By
$A_\alpha^{\alpha'}=\partial x^{\alpha'}/\partial x^\alpha$ and $|A| :=
\text{det}(A_{\alpha'}^{\alpha})$ we denote the Jacobian matrix and its determinant. To clarify the
following discussion, we introduce the convention that, if a product of $A_\alpha ^{\alpha'}$
occurs in a transformation formula, the kernel symbol~$A$ is written only once, e.g.
\begin{align}
A_\alpha ^{\alpha'}A_\beta ^{\beta'}A_\gamma ^{\gamma'} = A_{\alpha\; \beta\; \gamma}^{\alpha'
\beta' \gamma'}.
\end{align}
The electromagnetic fields $F_{\mu\nu}$ and $\mathcal{D}_{\mu\nu}$ in the primed and unprimed
coordinate systems are related by~\cite[Chap.~3.2]{post1997}:
\begin{align}
F_{\mu'\nu'} & = A_{\mu' \nu'}^{\mu\; \nu} F_{\mu\nu}, \nonumber\\
\mathcal{D}_{\mu'\nu'} & = |A|^{-1} A_{\mu' \nu'}^{\mu\; \nu}  \mathcal{D}_{\mu\nu}.
\label{eq:trafo_fields}
\end{align}
The Maxwell equations (\ref{eq:maxwell_kov1}) and (\ref{eq:maxwell_kov2}) have the same form in the primed coordinate system as in the unprimed system due to their natural
form-invariance\cite[Chap.~3.2]{post1997}, that is:
\begin{align}
\partial^{\mu'} F_{\nu'\sigma'} + \partial^{\sigma'}{F}_{\mu'\nu'}  + \partial^{\nu'} {F}_{\sigma'\mu'}  &= 0, \nonumber\\
\partial^{\mu'} \mathcal{D}_{\mu'\nu'} &= \frac{4\pi}{c} j_{\nu'}
\end{align}
with $j_{\nu'} = |A|^{-1} A_{\nu'}^{\nu} \,j_{\nu}$. In order to achieve form-invariance also for the
constitutive relation~(\ref{eq:matgleichung2}),
the material coefficients must transform as:
\begin{align}
\chi_{\mu'\nu'}^{\alpha_1'\beta_1'\cdots\alpha_n'\beta_n'} = |A|^{-1} \, A_{\mu' \nu'}^{\mu\; \nu}
\, A_{\alpha_1 \beta_1 \cdots \alpha_n \beta_n}^{\alpha_1' \beta_1' \cdots \alpha_n' \beta_n'} \,
\chi_{\mu\nu}^{\alpha_1\beta_1\cdots\alpha_n\beta_n} \label{eq:material_trafo}.
\end{align}
The proof of (\ref{eq:material_trafo}) is not difficult. To see this, we multiply both
sides of~(\ref{eq:matgleichung2}) by $|A|^{-1} \, A_{\mu' \nu'}^{\mu\; \nu}$, insert in the
right-hand side the identity (see~\cite{post1997}, Chap.~1.4)
\begin{align}
A_{\alpha_1 \beta_1 \cdots \alpha_n \beta_n}^{\alpha_1' \beta_1' \cdots \alpha_n' \beta_n'} 
A^{\alpha_1 \beta_1 \cdots \alpha_n \beta_n}_{\alpha_1' \beta_1' \cdots \alpha_n' \beta_n'} = A_{\alpha_1 \beta_1 \cdots \alpha_n \beta_n}^{\alpha_1 \beta_1 \cdots \alpha_n \beta_n} 
 = 1
\end{align}
and replace the unprimed quantities with the corresponding primed ones. This yields:
\begin{align}
\mathcal{D}_{\mu\nu} &= 4\pi\,\sum_{n\ge1} \chi_{\mu\nu}^{\alpha_1\beta_1\cdots\alpha_n\beta_n}F_{\alpha_1\beta_1}\dots F_{\alpha_n\beta_n}\nonumber\\[3mm]
\underbrace{|A|^{-1} \, A_{\mu' \nu'}^{\mu\; \nu}\mathcal{D}_{\mu\nu}}_{\mathcal{D}_{\mu'\nu'}} &=
4\pi\,\sum_{n\ge1} \underbrace{|A|^{-1} \, A_{\mu' \nu'}^{\mu\; \nu} A_{\alpha_1 \beta_1 \cdots \alpha_n
\beta_n}^{\alpha_1' \beta_1' \cdots \alpha_n' \beta_n'} \,
\chi_{\mu\nu}^{\alpha_1\beta_1\cdots\alpha_n\beta_n}}_{\chi_{\mu'\nu'}^{\alpha_1'\beta_1'\cdots\alpha_n'\beta_n'}}
\underbrace{A^{\alpha_1 \beta_1 \cdots \alpha_n \beta_n}_{\alpha_1' \beta_1' \cdots \alpha_n'
\beta_n'}
F_{\alpha_1\beta_1}\dots F_{\alpha_n\beta_n}}_{F_{\alpha_1'\beta_1'}\dots F_{\alpha_n'\beta_n'}} \nonumber\\
\mathcal{D}_{\mu'\nu'} &= 4\pi\,\sum_{n\ge1}
\chi_{\mu'\nu'}^{\alpha_1'\beta_1'\cdots\alpha_n'\beta_n'}F_{\alpha_1'\beta_1'}\dots
F_{\alpha_n'\beta_n'}.
\end{align}
Thus, the constitutive equation transforms covariantly as required. Relation~(\ref{eq:material_trafo}) represents the general transformation law for linear and nonlinear materials parameters (including complicated magneto-electric coupling terms) and is equally valid for arbitrary  space-time coordinate transformations. 
\subsection{Moving media}
A typical physical situation where coordinate transformations play a role occurs when a medium 
moves relative to the observer. In such cases, the transformation law allows the calculation of the material parameters in the reference frame of the observer if the material parameters are known in the rest frame of the medium. Consider, for a example, a linear, homogeneous medium that is isotropic in its rest frame with the constitutive equations:
\begin{align}
\boldsymbol{D} &= \epsilon \boldsymbol{E}, \nonumber\\
\boldsymbol{H} &=\mu^{-1} \boldsymbol{B}.
\end{align}
According to (\ref{eq:lineare_terme}), the covariant material tensor~$\chi_{\mu\nu}^{\sigma\kappa}$ of an isotropic medium is related to the material parameters $\epsilon$ and $\mu$ by the equations:
\begin{align}
\epsilon &= 8\pi \,\chi_{0i}^{0i}\nonumber\\
\mu^{-1} &= 8\pi \, \chi_{kl}^{kl}
\label{eq:eps_mu_isotropic_material}
\end{align}
(modulo interchange of indices, e.g.~$\epsilon = -8\pi \,\chi_{i0}^{0i}$, etc.) while $\chi_{\mu\nu}^{\sigma\kappa}$ is zero otherwise. 
By means of the transformation rule of (\ref{eq:material_trafo}), we can now calculate the material equation in a reference frame in which the medium moves uniformly, say along the $x$-direction, with constant velocity~$v$. The corresponding coordinate transformation is given by $x^{\alpha'}=A^{\alpha'}_{\alpha}x^{\alpha}$ where the Jacobi matrix~$A^{\alpha'}_{\alpha}$ describes a Lorentz boost in the minus $x$-direction:
\begin{align}
A^{\alpha'}_{\alpha}=
\begin{pmatrix}
\gamma & \gamma\beta & 0 & 0\\
\gamma\beta & \gamma & 0 & 0\\
0 & 0 & 1 & 0 \\
0 & 0 & 0 & 1
\end{pmatrix}, \quad |A|=1
\end{align}
with $\beta = v/c$ and $\gamma=(1-\beta^2)^{-1/2}$. The inverse of the Jacobi matrix~$A^{\alpha}_{\alpha'}$ is obtained by replacing $\beta$ by $-\beta$ (corresponding to a velocity reversal). The material equation in the frame in which the medium is moving is obtained by applying the transformation law~(\ref{eq:material_trafo}) to the material equation~(\ref{eq:linearterm}) (expressed in primed coordinates), that is
\begin{align}
\mathcal{D}_{\mu'\nu'} = 4\pi\chi_{\mu'\nu'}^{\sigma'\kappa'}\,F_{\sigma'\kappa'}
 = 4\pi A_{\mu'\nu'\sigma\;\kappa}^{\mu\; \nu\; \sigma'\kappa'}\chi_{\mu\nu}^{\sigma\kappa}\,F_{\sigma'\kappa'}.
\end{align}
After a straightforward summing over the indices and by using the relation between the field tensors $\mathcal{D}_{\mu'\nu'}$ and $F_{\sigma'\kappa'}$ and the $\boldsymbol{E}'$, $\boldsymbol{B}'$, $\boldsymbol{D}'$ and $\boldsymbol{H}'$ fields (see Eqs.~(\ref{eq:field_strength_tensor}) and~(\ref{eq:displacement_tensor})) as well as the relation between the material tensor $\chi_{\mu\nu}^{\sigma\kappa}$ and $\epsilon$ and $\mu$ given by (\ref{eq:eps_mu_isotropic_material}), this can be rewritten in the more intuitive form:
\begin{align}
\boldsymbol{D}' & = \epsilon \boldsymbol{E}' + \gamma^2 \left(\epsilon - \frac{1}{\mu}\right)\frac{\boldsymbol{v}}{c}\times\left(\boldsymbol{B}'-\frac{\boldsymbol{v}}{c}\times\boldsymbol{E}'\right),\nonumber \\
\boldsymbol{H}' & = \frac{ \boldsymbol{B}'}{\mu} + \gamma^2 \left(\epsilon - \frac{1}{\mu}\right)\frac{\boldsymbol{v}}{c}\times\left(\boldsymbol{E}'+\frac{\boldsymbol{v}}{c}\times\boldsymbol{B}'\right)
\end{align}
in agreement with the findings in~\cite{rousseaux2008}. Since these equations are form-invariant under spatial rotation of the coordinate system, they are equally valid for arbitrary directions of the medium's velocity~$\boldsymbol{v}$. Obviously, a linear medium that is isotropic at rest becomes bianisotropic if it is moved relative to the observer 
\cite{rousseaux2008,landau1984,cheng1968,thompson2011}.

The mixing of the electric and magnetic response in moving media is a general effect which can also be observed for the nonlinear material coefficients. Consider, for example, the quadratic term of the material tensor given by~$\chi^{\sigma\kappa\alpha\beta}_{\mu\nu}$. In the frame in which the medium is moving the tensor transforms as
\begin{align}
\chi^{\sigma'\kappa'\alpha'\beta'}_{\mu'\nu'} =  A^{\sigma'\kappa'\alpha'\beta'\mu\;\nu}_{\sigma\;\kappa\;\alpha\;\beta\;\mu'\nu'}\,\chi^{\sigma\kappa\alpha\beta}_{\mu\nu}.\label{eq:trafo_moving_nonlinear}
\end{align}
As before, we can evaluate the sum, collect corresponding terms and identify the new nonlinear coefficients. However, in this generality the final expressions get very complicated and unintuitive. 
Therefore, in order to illustrate only the principle of the mixing of higher-order susceptibilities in moving media, we restrict to the second-order contribution to the \textit{electric} polarization~$P_i^{(2)}=\chi_{0i}^{\sigma\kappa\alpha\beta}\,F_{\sigma\kappa}F_{\alpha\beta}$. As described in (\ref{eq:zweite_ordnung}), the relevant components of the material tensor can be further split in terms with $\chi^{0n0l}_{0i}$, $\chi^{0nkl}_{0i}$ and $\chi^{mnkl}_{0i}$ (modulo interchange of indices) to separately account for contributions that are proportional to $E_iE_j$, $E_i B_j$ and $B_i B_j$, respectively. If we now apply the transformation (\ref{eq:trafo_moving_nonlinear}) separately to these three coefficients, the result can be expressed in a compact matrix form:
\begin{align}
\begin{pmatrix}
\chi^{0'n'0'l'}_{0'i'}\\[2mm]
\chi^{0'n'k'l'}_{0'i'}\\[2mm]
\chi^{m'n'k'l'}_{0'i'}
\end{pmatrix}
= A^{0\;i\;n'l'}_{0'i'n\;l}
\begin{pmatrix}
4A^{0'0'}_{0\;0} & 4A^{0'0'}_{0\;k} & A^{0'0'}_{m\;k}\\[2mm]
4A^{0'k'}_{0\;0} & 4A^{0'k'}_{0\;k} & A^{0'k'}_{m\;k}\\[2mm]
4A^{m'k'}_{0\;0} & 4A^{m'k'}_{0\;k} & A^{m'k'}_{m\;k}
\end{pmatrix}
\begin{pmatrix}
\chi^{0n0l}_{0i}\\[2mm]
\chi^{0nkl}_{0i}\\[2mm]
\chi^{mnkl}_{0i}
\end{pmatrix} .
\end{align}
Obviously, the time-dependence~$A^{i'}_0 = c^{-1}\partial x^{i'}/\partial t\ne 0$ due to the movement of the medium implies nonzero off-diagonal elements which induce a mixing of nonlinear material coefficients of electric and magnetic nature. This means, for example, that a Pockels medium at rest ($\chi^{0n0l}_{0i}\ne 0$) can display a Faraday effect ($\chi^{0'n'k'l'}_{0'i'}\ne 0$) if it is moved relative to the observer and vice versa.

\subsection{Time-independent transformations}
We now focus on the special case of time-independent, spatial transformations, i.e.~transformations with
\begin{align}
t' & = t \nonumber\\
x' & = x'(x,y,z) \nonumber\\
y' & = y'(x,y,z) \nonumber\\
z' & = z'(x,y,z)
\end{align}
which are of particular interest for the design of TO devices with tailored optical properties. In this case, we have
\begin{align}
A^{\alpha'}_{0}=\frac{\partial x^{\alpha'}}{c\,\partial t} =
\delta^{\alpha'}_{0'}\quad\text{and}\quad A^{0'}_{\alpha}= \frac{c\,\partial t'}{\partial
x^{\alpha}}=\delta^{0}_{\alpha}. \label{eq:zeitunabh}
\end{align}
Consequently, if the term $A^{\alpha'}_{i}T_{\alpha'\beta'}$ occurs in an expression, we can
replace the four-index $\alpha'$ by the three-index $i'$ (which runs from 1 to 3) since $A^{0'}_{i}=0$.
For example, the permittivity transforms under these conditions as
\begin{align}
\epsilon^{j'}_{i'} &= 8\pi\chi^{0'j'}_{0'i'} \nonumber\\
& = 8\pi |A|^{-1}A^{\mu\;\nu\;0'j'}_{0'i'\alpha\;\beta} \,\chi^{\alpha\beta}_{\mu\nu}\nonumber\\
& = 8\pi |A|^{-1}\delta^{\mu}_0\,A^{i}_{i'}\,\delta^{0}_{\alpha}\,A^{j'}_{j}\,\chi^{\alpha j}_{\mu i}\nonumber\\
& = 8\pi |A|^{-1}A^{i\;j'}_{i'j}\,\chi^{0j}_{0i}\nonumber\\
& = |A|^{-1}A^{i\;j'}_{i'j} \epsilon^{j}_{i}. \label{eq:epsilon_trafo}
\end{align}
Accordingly, the permeability and magneto-electric coupling terms in (\ref{eq:lineare_terme})
transform as:
\begin{align}
\mu_{i'}^{j'} & = |A|^{-1}A^{i\;j'}_{i'j} \mu_i^j , \nonumber\\\
\xi_{i'}^{j'} &= |A|^{-1}A^{i\;j'}_{i'j}\xi_i^j, \nonumber\\\
\zeta_{i'}^{j'} &=|A|^{-1}A^{i\;j'}_{i'j} \zeta_i^j.
\end{align}
Note that for purely spatial transformations there is no magneto-electric cross-mixing.

By the general transformation law (\ref{eq:material_trafo}), we can now calculate the
transformation behavior of the nonlinear material coefficients under spatial coordinate
transformations. For instance, we exemplarily calculate the second order susceptibility of the
electric polarization~$a_i^{jk}=4\chi^{0j0k}_{0i}$ which describes nonlinear optical effects such as second-harmonic generation or three-wave-mixing.
With the help of (\ref{eq:material_trafo}), the
general transformation of the second order tensor $\chi_{\mu\nu}^{\sigma\kappa\alpha\beta}$ is
\begin{align}
\chi_{\mu'\nu'}^{\sigma'\kappa'\alpha'\beta'} = |A|^{-1} \, A_{\mu' \nu'\sigma\; \kappa\; \alpha\; \beta}^{\mu\; \nu\;\sigma' \kappa' \alpha' \beta'} \,
\chi_{\mu\nu}^{\sigma\kappa\alpha\beta}
\end{align}
and for the special case of a spatial coordinate transformation (i.e.~with the conditions given in (\ref{eq:zeitunabh})), the second order susceptibility transforms as
\begin{align}
a_{i'}^{j'k'} &=4\chi^{0'j'0'k'}_{0'i'}\nonumber\\
& = 4 |A|^{-1} \, A_{0' i'\sigma \;\kappa\; \alpha\; \beta}^{\mu\; \nu\;0' j' 0' k'}  \,
\chi_{\mu\nu}^{\sigma\kappa\alpha\beta} \nonumber\\\
& = 4 |A|^{-1} \, A_{i'j\;k}^{i\;j'k'} \,\chi_{0i}^{0j0k}\nonumber\\\
& = |A|^{-1} \, A_{i'j\;k}^{i\;j'k'}\,a_{i}^{jk} \label{eq:trafo_sec_order_suscept}
\end{align}
where we applied similar calculation steps as in (\ref{eq:epsilon_trafo}). This shows that as soon as
the relation between the material coefficient of interest (linear or nonlinear) and the general
covariant tensor~$\chi_{\mu\nu}^{\alpha_1\beta_1\cdots\alpha_n\beta_n}$ is found, the
transformation law immediately follows from~(\ref{eq:material_trafo}).
\section{Twisted, nonlinear field concentrator}
In the following, we demonstrate that nonlinear problems in complex media with sophisticated linear and nonlinear optical properties can take a much simpler form in an appropriately transformed space. 

To provide an illustrative example of this calculation method, we consider a nonlinear, inhomogeneous
material that is illuminated by a strong laser field. We assume that the polarization of the
fundamental wave and the nonlinearity of the material match the condition for second harmonic
generation (SHG) and, as a goal, we want to calculate the spatial field distribution of the SHG wave inside
the medium. If the material is highly inhomogeneous, both the fundamental and the SHG wave follow a
complicated, distorted trajectory through the medium which generally hampers 
a numerical calculation and reliable prediction
of the SHG progress. However, if it is possible to find a coordinate transformation such that the
wave propagates uniformly along straight lines in the new coordinate system, the wave equation can
be readily solved by an ordinary integration along the field lines. A subsequent
back-transformation then yields the SHG field in the physical space. In the following, we
demonstrate and explain this calculation method for a specific example.

As a hypothetic, inhomogeneous material, we consider the special case of a TO medium, i.e.~an
artificial material whose permeability and permittivity tensors were obtained by applying a
coordinate transformation to a homogeneous, isotropic space. We suppose that the optical properties
of the homogeneous space are similar to that of vacuum with a permittivity and permeability equal
(or close) to unity. 
In the following, the coordinates of the uniform, isotropic space are indicated by
unprimed indices while the coordinates of the physical space of the inhomogeneous material are indicated by primed indices.

\begin{figure}
\centering
\includegraphics[width=0.5\columnwidth]{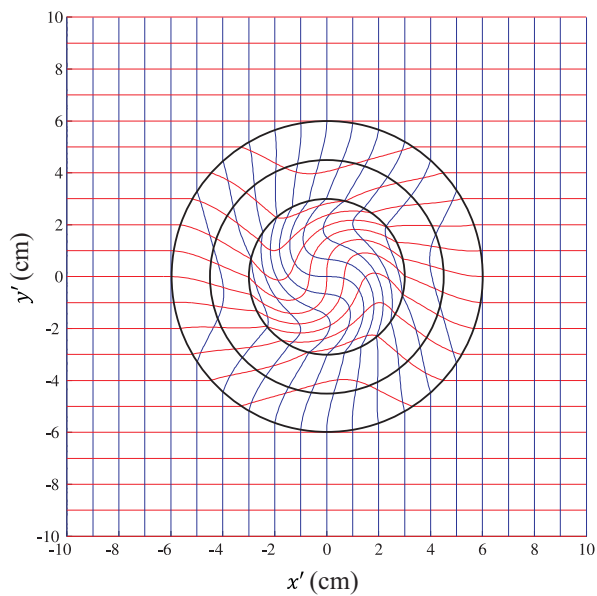}
\caption{Visualization of the twisted cylindrical concentrator. The mesh grid indicates lines with
constant values of $x$ and $y$ plotted in the coordinate system of $x'$ and $y'$. The radius of the
three displayed circles are $R_1=3$, $R_2=4.5$ and $R_3=6$.} \label{fig:mesh}
\end{figure}

For the transformation between the primed and unprimed system, we consider the following transformation (expressed in cylinder coordinates using $r^2=x^2+y^2$
and $\phi = \arctan(y/x)$):
\begin{align}
r'&=\begin{cases}
\frac{R_1}{R_2}r& \, 0\le r\le R_2\\[1mm]
\frac{R_3-R_1}{R_3-R_2}r-\frac{R_2-R_1}{R_3-R_2}R_3& \, R_2\le r\le R_3\\[1mm]
r & \,\text{otherwise}
\end{cases}\nonumber\\
\phi'&=\begin{cases}
\phi+ \frac{\pi}{2}\cos^2(\frac{\pi r}{2R_3})& \, 0\le r\le R_3\\[1mm]
\phi & \, \text{otherwise.}
\end{cases}
\label{eq:concentrator_trafo}
\end{align}
As illustrated in Fig.~\ref{fig:mesh}, the transformation compresses space of a
cylindrical region with radius~$R_2$ into a region with radius~$R_1$ at the cost of an expansion of
space between $R_1$ and $R_3$~\cite{rahm2008b}. In addition, the space experiences a radius-dependent twist about the $z$-axis~\cite{huanyangchen2007}. Note that the transformation addresses only the $x$- and $y$-coordinates while the $z$-coordinate
remains unchanged. For this reason, we can restrict the following discussion to the $xy$-submanifold of the space-time manifold.  The corresponding Jacobi matrix is:
\begin{align}
A_g &=\begin{pmatrix}
\frac{\partial r'}{\partial r} & \frac{\partial r'}{\partial \phi}  \\[1mm]
\frac{\partial \phi'}{\partial r} & \frac{\partial \phi'}{\partial \phi}
\end{pmatrix} 
=
\begin{cases}
\begin{pmatrix}
\frac{R_1}{R_2} & 0 \\[1mm]
-\frac{\pi^2}{4R_3}\sin(\frac{\pi r}{R_3}) & 1
\end{pmatrix}
& \, 0\le r\le R_2 \\[5mm]
\begin{pmatrix}
\frac{R_3-R_1}{R_3-R_2} & 0 \\[1mm]
-\frac{\pi^2}{4R_3}\sin(\frac{\pi r}{R_3}) & 1
\end{pmatrix}
& \, R_2\le r\le R_3 \\[5mm]
\text{diag}(1,1) &\,\text{otherwise.}
\end{cases}
\label{eq:concentrator_jacobi_cyl}
\end{align}
And the determinant of $A_g$ is:
\begin{align}
|A_g|=
\begin{cases}
\frac{R_1}{R_2} & \, 0\le r\le R_2 \\[1mm]
\frac{R_3-R_1}{R_3-R_2} & \, R_2\le r\le R_3\\[1mm]
1 &\,\text{otherwise.}
\end{cases}
\label{eq:concentrator_det_cyl}
\end{align}
The transformation (\ref{eq:concentrator_trafo}) is expressed in cylindrical coordinates.
For the calculation of the SHG, however, it is more advantageous to formulate the transformation in
Cartesian coordinates in the form $x'=x'(x,y)$ and $y'=y'(x,y)$. This is achieved by applying an
intermediate transformation $r = \sqrt{x^2+y^2}$ and $\phi = \arctan(y/x)$ in the unprimed system
and an inverse intermediate transformation $x' = r'\cos\phi'$ and $y' = r'\sin\phi'$ in the primed system. With these expressions, the transformation~(\ref{eq:concentrator_trafo}) can be
re-expressed in Cartesian coordinates by applying the following three subsequent transformations:
\begin{align}
(x,y)\stackrel{f}{\longrightarrow}(r,\phi)\stackrel{g}{\longrightarrow}(r',\phi')\stackrel{h}{\longrightarrow}(x',y').
\end{align}
According to the chain rule in higher dimension, the Jacobian matrix of a composite function is
just the product of the Jacobian matrices of the composed functions, that is
\begin{align}
A_{f\circ g\circ h} = A_fA_gA_h. \label{eq:concentrator_jacobi_cart}
\end{align}
As shown in the appendix, the Jacobian determinants of the intermediate transformations are
$|A_f|=1/r$ and $|A_h| = r'$, respectively. Consequently, the Jacobian determinant of the composite
is
\begin{align}
|A_{f\circ g\circ h}| = |A_f||A_g||A_h| = \frac{r'}{r}|A_g| \label{eq:concentrator_det_cart}
\end{align}
with the Jacobian determinant $|A_g|$ given in (\ref{eq:concentrator_det_cyl}). Once the
Jacobian matrix and its determinant are known, we can easily transform the fields and material
parameters from one coordinate system to the
other.\\

We now intend to calculate the SHG wave generated in the concentrator. For this purpose, we suppose that the material used for the construction of the concentrator exhibits a non-vanishing nonlinear susceptibility. To simplify matters, we further assume that this nonlinearity obeys the phase matching condition for frequency doubling if the fundamental wave and the second harmonic wave are both polarized in the $z$-direction.
In our notation, the
corresponding tensor component of the nonlinear susceptibility is
$a':=4\chi_{0'3'}^{0'3'0'3'}$ (see (\ref{eq:zweite_ordnung})). 
For the spatial distribution of~$a'$, we suppose
that the nonlinearity is only present in the inner cylindrical region of the material according to:
\begin{align}
a'(r')=
\begin{cases}
a_0 & \, 0\le r'\le R_1 \\[1mm]
0 &\,\text{otherwise.}
\end{cases}
\label{eq:concentrator_nonlin_primed}
\end{align}
This could, for example, be realized by doping the center of the concentrator with some nonlinear material.

As fundamental
wave we assume an incident monochromatic plane wave that initially propagates along the $x$-axis while the electric field
vector is polarized in the $z$-direction. In the physical space of the inhomogeneous material (spanned by primed coordinates), the fundamental wave takes the form:
\begin{align}
E'_\omega(x',y',t) = \mathcal{E}'_\omega\,e^{i(k_{\omega\,x'}'x'+k_{\omega\,y'}'y'-\omega t)}
\end{align}
where $\mathcal{E}'_{\omega}$ denotes the amplitude of the wave and $\boldsymbol{k}'_\omega= (k'_{\omega\,x'},k'_{\omega\,y'})$ is the wave vector in the medium. 
For the second harmonic wave, we represent the electric field by
\begin{align}
E'_{2\omega}(x',y',t) =
\mathcal{E}'_{2\omega}(x',y')\,e^{i(k'_{2\omega\,x'}x'+k'_{2\omega\,y'}y'-2\omega t)}
\end{align}
with the SHG wave vector $\boldsymbol{k}'_{2\omega}=2\boldsymbol{k}'_{\omega}$ (phase matched case).
In the slowly varying amplitude approximation for the second harmonic wave and the undepleted pump
approximation (i.e.~$\mathcal{E}'_\omega$ is constant), the wave equation for the SHG
amplitude~$\mathcal{E}'_{2\omega}(x',y')$ is given by~\cite{boyd2008}:
\begin{align}
\left(k'_{2\omega\,x'}\frac{\partial}{\partial x'} + k'_{2\omega\,y'}\frac{\partial}{\partial
y'}\right)\mathcal{E}'_{2\omega}(x',y') = \kappa \mathcal{E}_\omega^{'2} a'(x',y')
\label{eq:nonlinear_waveequation}
\end{align}
with $\kappa = -16\pi i \omega^2/c^2$. This is a partial differential equation in two dimensions
for the second harmonic field amplitude $\mathcal{E}'_{2\omega}(x',y')$ which is very
difficult to solve in general. However, the complexity can be significantly reduced if we apply a coordinate
transformation to the uniform space (spanned by unprimed coordinates).
\begin{figure}
\centering
\begin{tabular}{c}
\includegraphics[width=0.85\columnwidth]{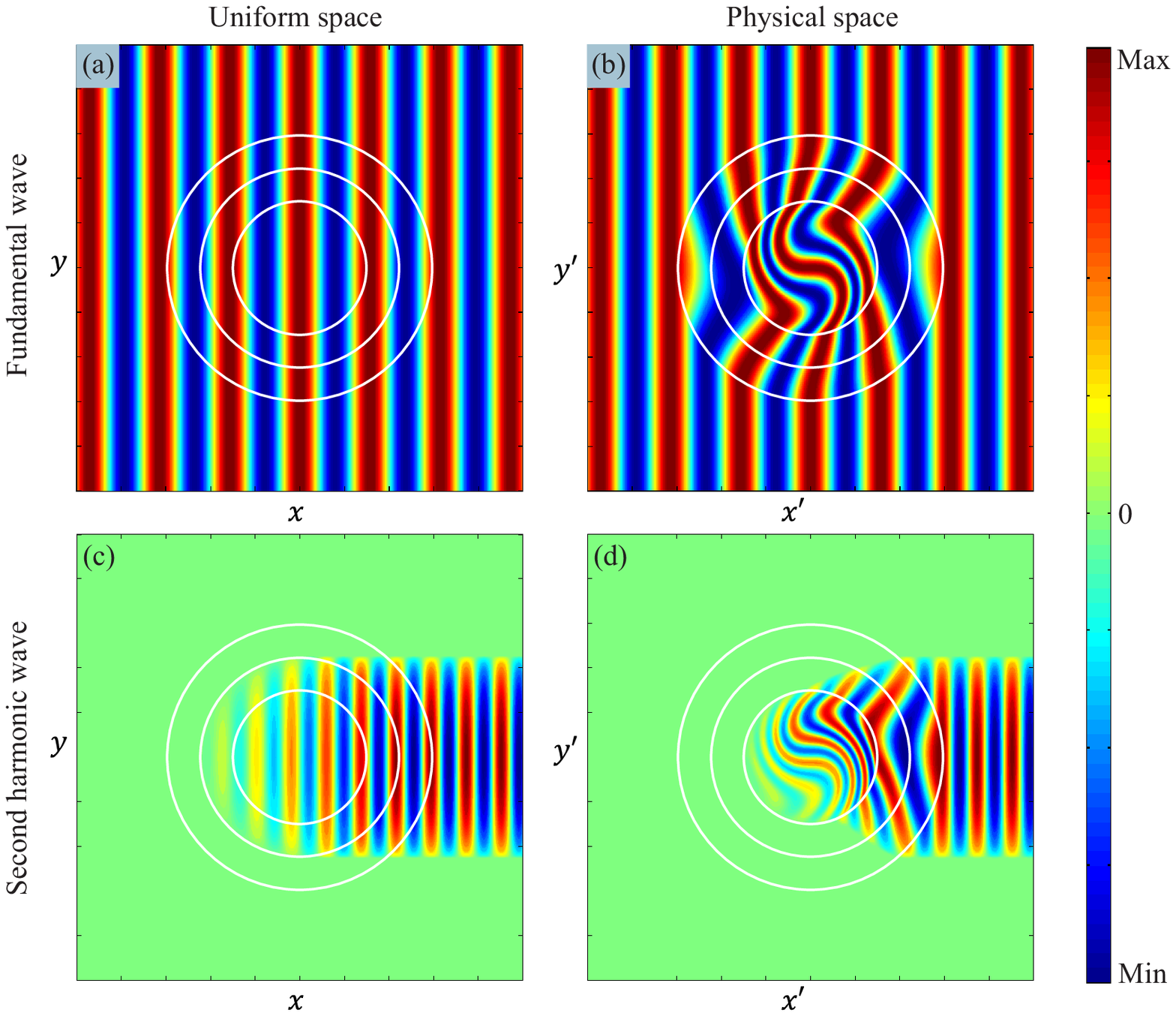}
\end{tabular}
\caption{Second harmonic wave generation in a nonlinear cylindrical concentrator. (a)~Electric field of the fundamental wave in the uniform space (unprimed system) and (b)~in the inhomogeneous, physical space (primed system).
(c)~Electric field of the second harmonic wave in the unprimed system and (d)~in the primed system. The radius of the three
displayed circles are $R_1=3$, $R_2=4.5$ and $R_3=6$.} \label{fig:overview}
\end{figure}

According to (\ref{eq:trafo_fields}) and the Jacobian matrix of the transformation given by
(\ref{eq:concentrator_jacobi_cart}) (with suppressed subscript $f\circ g\circ h$), the
$z$-component of the electric field transforms as
\begin{align}
E'_{z'} = F_{0'3'} = A_{0' 3'}^{\mu\; \nu} F_{\mu\nu}  = F_{03} = E_z
\end{align}
since $t'=t$ and $z'=z$ implies $A_{0' 3'}^{\mu\; \nu} = \delta_{0}^{\mu} \delta_{3}^{\nu}$. Consequently, the electric fields of both the fundamental and SHG wave remain
unchanged under coordinate transformation (but the functional dependence on the underlying coordinate system
may be different). Since the unprimed system is homogeneous and isotropic with a refractive index
equal to one, the electric field of the fundamental wave in the unprimed system has the form of a
uniform plane wave propagating in the $x$-direction:
\begin{align}
E_\omega(x,t) = \mathcal{E}_\omega\,e^{i(k_{\omega\,x}x-\omega t)}
\label{eq:fund_wave_unprimed}
\end{align}
with constant amplitude~$\mathcal{E}_\omega=\mathcal{E}'_\omega$ and $k_{\omega\,x}=\omega/c$. 
From
this expression, the electric field in the primed coordinate system is readily obtained
by expressing~$x$ 
by $x=x(x',y')$.
The real part of the fundamental wave in the primed and unprimed system is plotted in
Figs.~\ref{fig:overview}(a) and \ref{fig:overview}(b), respectively.

Since the fundamental wave propagates straightly along the $x$-direction in the unprimed system and
since we assumed phase matching for the SHG process, it follows that the wave vector of the SHG
wave is given by $k_{2\omega\,x}=2k_{\omega\,x}=2\omega/c$ and $k_{2\omega\,y}=0$ (on the
considered length scale, beam divergence due to diffraction can be neglected). Consequently, the
wave equation~(\ref{eq:nonlinear_waveequation}) for the SHG process reduces in the unprimed system
to the simple expression
\begin{align}
\frac{\partial}{\partial x} \mathcal{E}_{2\omega}(x,y) = \frac{\kappa c}{2\omega}
\mathcal{E}_\omega^{2} a(x,y)
\end{align}
which can be immediately integrated to:
\begin{align}
\mathcal{E}_{2\omega}(x,y) = \frac{\kappa c}{2\omega} \mathcal{E}_\omega^{2} \int_{-\infty}^x
\!ds \,a(s,y). \label{eq:integral}
\end{align}
The remaining unknown is the nonlinearity~$a=4\chi^{0303}_{03}$, i.e.~the relevant tensor component for the SHG process in the unprimed system. 
According to the transformation law derived in (\ref{eq:trafo_sec_order_suscept}), the nonlinearity transforms as:
\begin{align}
 a' &= 4\chi^{0'3'0'3'}_{0'3'} 
 =4 |A|^{-1}A^{i\;3'3'}_{3'j\,k}\chi^{0j0k}_{0i}
 =4|A|^{-1} \chi^{0303}_{03}
 =|A|^{-1} a
\end{align}
since $z'=z$ implies that $A_{3'j\;k}^{i\;3'3'} = \delta_{3}^{i}\delta_{j}^{3}\delta_{k}^{3}$.
With the determinant $|A|$ given by
(\ref{eq:concentrator_det_cart}) and (\ref{eq:concentrator_det_cyl}), the nonlinearity
$a'$ defined in (\ref{eq:concentrator_nonlin_primed}) and the relation between $r$
and $r'$ given by (\ref{eq:concentrator_trafo}), we can calculate the nonlinearity in the
unprimed system to be:
\begin{align}
a=\frac{r'}{r}|A_g| a' =
\begin{cases}
\left(\frac{R_1}{R_2}\right)^2 a_0 & \, 0\le r\le R_2 \\[1mm]
0 &\,\text{otherwise.}
\end{cases}
\end{align}
As expected, the nonlinearity in the unprimed system is reduced by a factor of $(R_1/R_2)^2$
compared to the nonlinearity in the primed system (see (\ref{eq:concentrator_nonlin_primed}))
because the cylindrical region in which the nonlinear substance is located experiences a space
expansion from radius $R_1$ to $R_2$ if we perform a coordinate transformation from the primed to
the unprimed system.

Now we can (even analytically) evaluate the integral~(\ref{eq:integral}) for the amplitude
$\mathcal{E}_{2\omega}(x,y)$ of the SHG wave. A subsequent multiplication with the propagation
phasor yields the SHG wave in the unprimed system in the form:
\begin{align}
E_{2\omega}(x,y,t) = \mathcal{E}_{2\omega}(x,y)\,e^{2i\omega(x/c- t)}.
\end{align}
By applying the inverse transformation (i.e.~expressing $x$ and $y$ by $x=x(x',y')$ and
$y=y(x',y')$), we finally obtain the SHG wave in the physical space of the primed coordinates. The real part of the resulting
SHG field distribution in the two coordinate systems is plotted in Figs.~\ref{fig:overview}(c) and
\ref{fig:overview}(d), respectively.

The proposed twisted nonlinear concentrator is certainly a somewhat constructed example since the exploited uniform space was already presumed in the design of the concentrator.
However, the decisive step in the calculation---the straightening of the wave trajectories inside the medium by applying an appropriate coordinate transformation---is in principle always possible in any inhomogeneous media with continuously varying material properties. The proposed technique of simplifying nonlinear processes in inhomogeneous media is therefore not restricted to the presented example, but covers a wide application range.

\section{Conclusion}
We proposed a theoretical framework for the incorporation of nonlinear effects within the concept of transformation optics (TO). In this context, we derived a general expression for the calculation of linear and nonlinear electromagnetic material parameters under arbitrary coordinate transformations. The transformation law is formulated in a manifestly covariant form that allows the simultaneous treatment of electric, magnetic and magneto-electric cross-coupling terms and is applicable to both temporal and spatial transformations.

As a first application example, we calculated the linear and nonlinear constitutive material relations in a moving medium.
As expected from the transformation of electromagnetic fields, the movement of the medium leads to a mixing of the electric and magnetic response, where the obtained expressions for the linear constitutive equations were in agreement with the relations given in the literature. 
In addition, we showed that the movement of the medium also implies a 
mixing of the nonlinear material properties which describe cross-coupled nonlinear interactions between the $\boldsymbol{E}$- and $\boldsymbol{B}$-fields and the medium. 
This means, for example, that a Pockels medium at rest can display a Faraday effect if the material is moved relative to the observer, and vice versa.

In the final part of the paper we focused on time-independent, spatial coordinate transformations which are of particular interest for the design of nonlinear TO devices. 
As an illustrative example of such a device, we presented a 
twisted nonlinear field concentrator and calculated the second harmonic wave that is generated when the concentrator is illuminated by a strong laser field. In this respect, we demonstrated that 
sophisticated nonlinear phenomena in complex media can take a much simpler form if an appropriate coordinate transformation is applied. 

The considerations have shown that the incorporation of nonlinear susceptibilities in the TO approach provides a promising computation method for calculating nonlinear effects in moving or inhomogeneous media and offers new opportunities for the design of novel optical devices with tailored nonlinear properties.

\section*{Appendix}
\subsection*{Summary of the linear terms in equation~(\ref{eq:matgleichung2})} The linear terms
in (\ref{eq:matgleichung2}) (the free-space contribution and the linear response of the material)
can be written as
\begin{align}
\mathcal{D}_{\mu\nu} &= F_{\mu\nu} + 4\pi \chi_{\mu\nu}^{\sigma\kappa}F_{\sigma\kappa}\nonumber\\
& = \frac{1}{2}\left(F_{\mu\nu}-F_{\nu\mu}\right) + 4\pi\chi_{\mu\nu}^{\sigma\kappa}F_{\sigma\kappa}\nonumber\\
& = \frac{1}{2}\left(\delta_{\mu\nu}^{\sigma\kappa}-\delta_{\nu\mu}^{\sigma\kappa}\right) F_{\sigma\kappa}+ 4\pi\chi_{\mu\nu}^{\sigma\kappa}F_{\sigma\kappa}\nonumber\\
& = \left(\frac{1}{2}\left(\delta_{\mu\nu}^{\sigma\kappa}-\delta_{\nu\mu}^{\sigma\kappa}\right)
+4\pi\chi_{\mu\nu}^{\sigma\kappa} \right)F_{\sigma\kappa}.
\end{align}
And finally, the last line of (\ref{eq:matgleichung2}) follows after the re-definition
$4\pi\chi_{\mu\nu}^{\sigma\kappa}\,(\text{new}):=\frac{1}{2}\left(\delta_{\mu\nu}^{\sigma\kappa}-\delta_{\nu\mu}^{\sigma\kappa}\right)
+4\pi\chi_{\mu\nu}^{\sigma\kappa}$.\\

\subsection*{Proof of equation~(\ref{eq:lineare_terme})}
By using the identity
\begin{align}
F_{ij} & = \frac{1}{2} \left(F_{ij}-F_{ji}\right) =
\frac{1}{2}\left(\delta_{ij}^{mn}-\delta_{ji}^{mn}\right) F_{mn} \nonumber\\
&= \frac{1}{2}g^k_{ij}g^{mn}_k F_{mn} = - g^k_{ij} B_k
\end{align}
the components of the electric displacement field are
\begin{align}
D_i&=\mathcal{D}_{0i} =4\pi\chi_{0i}^{\sigma\kappa}\,F_{\sigma\kappa} \nonumber\\
& = 8\pi \chi_{0i}^{0j} F_{0j} + 4\pi \chi_{0i}^{mn}F_{mn} \nonumber\\
& = 8\pi \chi_{0i}^{0j} E_j  - 4\pi g^j_{mn} \chi_{0i}^{mn} B_j
\end{align}
and, accordingly, the components of the magnetic fields are given by
\begin{align}
H_i &= -\frac{1}{2}g_i^{mn}\mathcal{D}_{mn} = -2\pi g_i^{mn}\chi_{mn}^{\sigma\kappa}\,F_{\sigma\kappa}\nonumber \\
& = -4\pi g_i^{mn} \chi_{mn}^{0j} F_{0j} - 2\pi g_i^{mn} \chi_{mn}^{kl} F_{kl} \nonumber\\
& = -4\pi g_i^{mn} \chi_{mn}^{0j} E_j + 2\pi g_i^{mn} g^j_{kl} \chi_{mn}^{kl} B_j.
\end{align}
By comparing the expressions of $D_i$ and $H_i$ with
\begin{align}
D_i &= \epsilon_i^jE_j + \xi_i^jB_j \nonumber\\
H_i &= \zeta_i^jE_j + (\mu^{-1})_i^jB_j,
\end{align}
we obtain the relations given in (\ref{eq:lineare_terme}).

\subsection*{Number of independent components in $\chi_{\mu\nu}^{\alpha_1\beta_1\cdots\alpha_n\beta_n}$}
Due to the symmetry properties 
\begin{align}
\chi_{\mu\nu}^{\alpha_1\beta_1\alpha_2\beta_2} = -\chi_{\nu\mu}^{\alpha_1\beta_1\alpha_2\beta_2} =
\chi_{\mu\nu}^{\alpha_2\beta_2\alpha_1\beta_1} = -\chi_{\mu\nu}^{\beta_1\alpha1\alpha_2\beta_2}
\end{align}
(exemplary for $n=2$), there are only 6 independent combinations for each upper index pair $\alpha_i\beta_i$, e.g.~$\alpha_i\beta_i=01,\,02,\,03,\, 12,\,13,\,23$. Furthermore, in the sequence $\alpha_1\beta_1\cdots\alpha_n\beta_n$ all $n$~pairs $(\alpha_i\beta_i)$ can be permuted as a whole without changing the value of $\chi_{\mu\nu}^{\alpha_1\beta_1\cdots\alpha_n\beta_n}$. Consequently, for fixed lower index pair $\mu\nu$, the number~$N$ of independent tensor components is equal to the number 
of combinations of $n$~labels where each label can take 6~values (with repetition). This number is given by:
\begin{align}
N(n) =
\begin{pmatrix}
6+n-1 \\ n
\end{pmatrix} = \frac{(5+n)!}{5!n!}.
\end{align}
For instance, $N(1)=6$, $N(2)=21$, $N(2)=56$, etc. 
Finally, we can also vary the lower index pair~$\mu\nu$ for which also 6 independent combinations are possible due to the above symmetries. Hence, the total number of independent components of the tensor 
$\chi_{\mu\nu}^{\alpha_1\beta_1\cdots\alpha_n\beta_n}$ is given by $6N(n)$.

\subsection*{Intermediate transformations used in equation~(\ref{eq:concentrator_det_cart})}
The first intermediate transformation~$f$ is given by
\begin{align}
r & = \sqrt{x^2+y^2}\nonumber\\
\phi &= \arctan(y/x)
\end{align}
with the corresponding Jacobian matrix and determinant:
\begin{align}
A_f =
\begin{pmatrix}
\frac{\partial r}{\partial x} & \frac{\partial r}{\partial y}  \\[1mm]
\frac{\partial \phi}{\partial x} & \frac{\partial \phi}{\partial y}
\end{pmatrix}
=
\begin{pmatrix}
\frac{x}{r} & \frac{y}{r} \\[1mm]
-\frac{y}{r^2} & \frac{x}{r^2}
\end{pmatrix},
\quad |A_g| = \frac{1}{r}.
\end{align}
The inverse intermediate transformation is given by:
\begin{align}
x' & = r'\cos\phi'\nonumber\\
y' &= r'\sin\phi'
\end{align}
The corresponding Jacobian matrix and determinant are:
\begin{align}
A_h =
\begin{pmatrix}
\frac{\partial x'}{\partial r'} & \frac{\partial x'}{\partial \phi'}  \\[1mm]
\frac{\partial y'}{\partial r'} & \frac{\partial y'}{\partial \phi'}
\end{pmatrix}
=
\begin{pmatrix}
\cos\phi' & -r'\sin\phi' \\[1mm]
\sin\phi' & r'\cos\phi'
\end{pmatrix},
\quad |A_h| = r'
\end{align}


\section*{References}
\providecommand{\newblock}{}

\end{document}